\documentstyle[12pt,aaspp4]{article}
\begin{document}
\newcommand{\up}[1]{\ifmmode^{\rm #1}\else$^{\rm #1}$\fi}
\newcommand{\zdot}{\makebox[0pt][l]{.}}
\newcommand{\upd}{\up{d}}
\newcommand{\uph}{\up{h}}
\newcommand{\upm}{\up{m}}
\newcommand{\ups}{\up{s}}
\newcommand{\arcd}{\ifmmode^{\circ}\else$^{\circ}$\fi}
\newcommand{\arcm}{\ifmmode{'}\else$'$\fi}
\newcommand{\arcs}{\ifmmode{''}\else$''$\fi}

\title{The Araucaria Project. Population effects on the V and I band 
magnitudes of red clump stars
}
\author{Grzegorz Pietrzy{\'n}ski}
\affil{Universidad de Concepci{\'o}n, Departamento de Astronomia,
Casilla 160-C,
Concepci{\'o}n, Chile}
\affil{Warsaw University Observatory, Al. Ujazdowskie 4, 00-478, Warsaw,
Poland}
\authoremail{pietrzyn@astrouw.edu.pl}
\author{Marek G{\'o}rski}
\affil{Warsaw University Observatory, Al. Ujazdowskie 4, 00-478, Warsaw,
Poland}
\authoremail{mgorski@astrouw.edu.pl}
\author{Wolfgang Gieren}
\affil{Universidad de Concepci{\'o}n, Departamento de Astronomia,
Casilla 160-C, Concepci{\'o}n, Chile}
\authoremail{wgieren@astro-udec.cl}
\author{David, Laney}
\affil{Department of Physics and Astronomy, N283 ESC, Brigham Young University, Provo, 
UT}
\authoremail{cdl@saao.ac.za}
\author{Andrzej Udalski}
\affil{Warsaw University Observatory, Al. Ujazdowskie 4, 00-478, Warsaw,
Poland}
\authoremail{udalski@astrouw.edu.pl}
\author{Anna Ciechanowska}
\affil{Warsaw University Observatory, Al. Ujazdowskie 4, 00-478, Warsaw,
Poland}
\authoremail{aciechan@astrouw.edu.pl}

\begin{abstract}
We present measurements of the V and I band magnitudes of red clump stars 
in 15 nearby galaxies obtained from recently published homogenous HST photometry.
Supplementing these results  with similar data for another 8 galaxies 
available in the literature the populational effects on the V and I band magnitudes of 
red clump stars were investigated. Comparing red clump magnitudes 
with the I-band magnitude of the TRGB in a total sample of 23
galaxies possessing very different environments we demonstrate that 
population effects strongly affect both the V and I band magnitude of red 
clump stars in a complex way. Our empirical results basically confirm the 
theoretical results of Girardi and Salaris, and show that 
optical (VI) photometry of red clump stars is not an accurate method
for the determination of distances to nearby galaxies at the present 
moment, as long as the population effects are not better calibrated, both
empirically and theoretically. Near infrared photometry is a much better way to
measure galaxy distances with red clump stars given its smaller sensitivity
to population effects. 
\end{abstract}

\keywords{distance scale - galaxies: distances and redshifts}

\section{Introduction}
The main goal of our long-term Araucaria Project
(Gieren et al. 2005) is to substantially improve the extragalactic distance 
scale calibration with observations of several major distance indicators 
in a large sample of nearby galaxies possessing markedly different environments. 
Using high-quality optical and near-infrared photometry, and spectroscopic data 
collected for 
Cepheids, TRGB, RR Lyrae stars, blue supergiants, eclipsing binaries and red clump stars
we are carefully addressing two of the most important sources of systematic error 
associated with the whole procedure of calibrating the cosmic 
distance scale: reddening, and the effect of population (metallicities and ages)
on the various stellar distance indicators scrutinized in our program. In addition,
this work is expected to produce a truly accurate (1 \% total error) distance 
determination to the LMC which is the fiducial galaxy to which most measures
of distances to other, more distant galaxies are currently tied. 

The helium-burning stars of intermediate age called red clump stars are in principle a
very attractive tool for distance determination, given their small ranges
in absolute magnitude in any given photometric passband, their presence in
most galaxies in large numbers which allows to determine the mean red clump magnitude
typically to 0.01 mag or better, 
and the fact that red clump stars are very numerous also in the 
solar neighborhood. About 1000 red clump stars have trigonometric 
parallaxes measured with an accuracy better than 10 \% by the Hipparcos satellite,
which made it possible to calibrate the absolute magnitudes of the red clump with high 
accuracy (Paczynski and Stanek 1998, Alves 2000, 
Laney and Pietrzynski 2009). 

The first measurements of the distance to the LMC based on the mean I-band magnitude
of red clump stars led to a very short distance modulus of 18.23 mag 
(Udalski et al. 1998, Udalski  2000) and triggered a lively dispute in the 
astronomical community about the possible reality of a "short distance scale". 
Subsequent observations of red clump stars in clusters in 
the Magellanic Clouds indicated that the dependence of the mean I-band 
magnitude on age is practically negligible (within ~ 0.05 mag) for the age 
range 2-8 Gyr (Udalski 1998). However, these results had to be taken with some caution
given the fact that the studied LMC and SMC clusters had slightly 
different distances. Sarejedini (1999) using data of Galactic open 
clusters found that there are important population effects on both 
V and I band magnitudes of red clump stars. 
Detailed studies of the dependence of ${\rm I}_{\rm RC}$
on metallicity were performed  on a large set of nearby red clump 
stars with accurate distances and spectroscopic metallicities (McWilliam 1990). 
Within the range of metallicities covered by these nearby red clump stars  
(-0.6  $<$ [Fe/H] $<$ +0.35; McWilliam 1990, Liu et al. 2007) the 
slope of the brightness-metallicity relation 
was found to be a rather modest 0.14 mag/dex (Udalski 2000, ApJ, 531, L25). 

Important progress was made by Alves (2000) who took the work to the near-infrared
domain and calibrated the K-band absolute magnitudes of a large sample of 
Hipparcos-observed nearby red clump stars. An outstanding and obvious 
advantage of using K-band data is that it practically eliminates any dependence 
of the obtained results on the adopted reddening. Alves (2000) was also able to show
that the mean K-band magnitude of the red clump does not depend on 
metallicity in the range from -0.5 to 0.0 dex.  The distance modulus 
measured to the LMC using the Alves calibration surprisingly and consistently 
turned out to be 
close to 18.50 mag (Alves et al. 2002, Pietrzynski and Gieren 2002), in conflict
with the earlier result from the I-band magnitude of the red clump in the LMC. 
Girardi and Salaris (2001), and  Salaris and Girardi (2002), assuming a star 
formation history and chemical evolution history, computed theoretical population  
corrections which must be applied to the optical and near-IR magnitudes 
of red clump stars in order to calculate correct distances to nearby galaxies. 
They argued that the observed discrepancy between the distances to the LMC measured
from the red clump in I and K bands is a consequence of the larger population effect
in the I band. Unfortunately, the corrections predicted by their models are very 
uncertain (about 0.1 mag) due to uncertain star formation and chemical evolution
histories and shortcomings in the stellar models,
making it imperative to improve the accuracy of these
theoretical predictions, or providing an accurate empirical determination of
the population effect in the different bands.

In the course of the Araucaria project we determined the mean near-IR magnitudes of 
the red clump in four Local Group galaxies (LMC, SMC, Carina and 
Fornax; Pietrzynski, Gieren and Udalski 2003). Comparison of the measured 
mean K-band magnitude to the magnitudes of other distance indicators 
(Cepheids, TRGB, RR Lyrae stars) showed, in very good agreement with the results
 of Alves (2000), and of van Helshoecht and Groenewegen (2007),  that in this band the 
population effects on the brightness of the red clump are indeed very small.  
Again, the distance moduli based on the near-IR calibration of the red clump method
to the four studied galaxies turned to be longer by almost exactly 
0.21 mag when comparing to the corresponding distances obtained from the 
optical I-band photometry. Therefore the large discrepancy between the distances
obtained in the optical and near-IR 
may be either explained by very significant population effects on the optical 
brightness of the red clump stars, or by a large zero point error in one of the 
calibrations. Since in particular the calibration of Alves (2000) is based 
on very old photometry collected in 1960s, a significant zero point error
in the calibration did not seem to be completely ruled out. 
In order to check on this possibility very recently a new, modern calibration of 
the K, H and J band magnitudes of red clump stars has been obtained (Laney and Pietrzynski 
2009) which is based on observations of nearby Galactic red clump  stars with the 
infrared photometer attached to the 0.75-m telescope at the 
SAAO observatory. With the  new 
accurate calibration (${\rm M}^{K}_{\rm RC} = -1.615 \pm 0.03$ mag)
we measured a LMC distance of 18.47 $\pm$ 0.03 mag 
in near-perfect agreement with our our recent distance determination to this 
galaxy based on a late - type eclipsing binary 
(Pietrzynski et al. 2009) and near-infrared photometry of 
RR Lyrae stars (Szewczyk et al. 2009). All these results indicated that there
are indeed large population effects on the optical magnitudes of red clump stars,
as oppose to their near-infrared magnitudes, but a convincing empirical demonstration 
of this conjecture had yet to be made, and is the purpose of this paper where we
take advantage of the recently published (Jacobs et al. 2009) uniform HST photometry
of a significant number of nearby galaxies containing large numbers of red clump
stars and stars populating the red giant branch.

\section{Mean V and I band red clump magnitudes}
The relatively faint absolute magnitudes of red clump stars 
seriously limited the number of nearby galaxies in which
accurate mean V and I band magnitudes of the red clump were measured and 
analyzed so far (e.g. Udalski et al. 2000), preventing sound
empirical conclusions about the population effects in these bands
on the brightness of these stars.
Very recently Jacobs et al. (2009) published a compilation of 
deep HST photometry of some 250 galaxies in the Local Volume. 
Taking advantage of this new and formidable database we decided to 
use this deep and uniform photometry to calculate  accurate 
V and I band mean red clump magnitudes in 15 Local Group galaxies in which
the HST photometry reaches at least 
two magnitudes below the faintest red clump stars, making a very precise
measurement of the clump magnitude possible. We did this measurement applying
fitting function (1), which consists of a Gaussian component representing the 
red clump stars and a second-order polynomial approximating the stellar background
(RGB stars), to the data. Based on this procedure the mean V and I band magnitudes of the
red clump were determined in each target galaxy with a typical accuracy of 0.01 mag  
(see Figure 1, and Table 1).

$$n(K)=a+b(K-K^{\rm max})+c(K-K^{\rm max})^2 +
 \frac{N_{RC}}{\sigma_{\rm RC}\sqrt{2\pi}} \exp\left[-\frac{(K-K^{\rm
max})^2}{2\sigma^2_{\rm RC}}\right]\eqno(1)$$ \\

In order to determine a uniform set of metallicities for our target galaxies,
 we applied the calibration of
the V-I color of the Red Giant Branch (RGB) versus metallicity relation 
of  Da Costa and Armandroff (1990) to the same HST data. 
We note that this relation was derived 
from old globular clusters and might need a correction for galaxies 
having a significant population of younger RGB stars. In order to roughly 
estimate the expected change in the metallicity determination from the
Da Costa and Armandroff relation for such systems
(e.g. NGC 6822, IC 1613, M33, M31) we 
adopted several star formation histories and computed the
expected corrections. These were found to be
very small (about 0.1 dex) which is likely within the systematic uncertainties
of the metallicities derived from the Da Costa and Armandroff relation.
Therefore we decided not to apply them to the calculated metallicities..
The results are presented in Table 1, together with 
the values of the  I-band magnitude of the TRGB measured from 
the same data by Rizzi et al. (2008). Foreground reddenings 
estimated from the reddening maps of Schlegel et al. (1999) are also given.

\section{Comparison with the I band of TRGB}
In this section we will determine the differences between the mean I-band 
magnitudes of the red clump and the corresponding I-band magnitudes of the TRGB
in our sample of galaxies, and relate these magnitude differences to the metallicities
in Table 1.
For several reasons such a differential comparison provides a very strong test on 
populational effects on the brightness of the red clump. Indeed, many 
nearby galaxies possess large populations of RGB and red clump stars, so the corresponding 
brightness of these two distance indicators can be measured with very good 
accuracy. Moreover, reasonably assuming that both red clump star 
and TRGB magnitudes are affected by interstellar extinction in the same way, 
the  differential I-band magnitudes will be free of problems 
related to the reddening, and potential errors of the photometric zero 
points. Using the data presented in Table 1 together with the corresponding
information for 8 additional galaxies tabulated by Udalski (2000), we calculated 
the ${\rm I}_{\rm TRGB} - {\rm I}_{\rm RC}$ for 23 nearby galaxies  and plotted these
differences versus the corresponding metallicities in Fig. 2. Since it is widely accepted 
now that 
the absolute I-band magnitude of the TRGB depends only very weakly on metalliciy and age
(e.g. some 0.04 mag/dex, Rizzi et al. 2007) the scatter on  Fig. 2
must be mostly related to the populational dependence of the mean I-band magnitude of 
the red clump stars.

\section{Discussion}
Several features which are seen on Fig. 2 deserve to be commented.
First of all, a scatter among the data points up to 0.4 mag is present, which 
fully confirms the results obtained from the theoretical models 
of Girardi and Salaris (1999) that there is a very significant population effect on 
the I-band magnitude of red clump stars. Fig. 2 shows this fact for the first time
empirically, in a very clear fashion. Furthermore, a trend is present in the sense that
the red clump in I becomes fainter with respect to the TRGB in I as the stellar
metallicity increases.

The mean I-band 
magnitude of red clump stars calculated from a mixture of different populations 
can in principle depend on metallicity in a very complicated way through the star 
formation and chemical evolution histories. The simple linear relation 
in Fig. 2 was plotted to demonstrate that a strong population effect 
is present, rather than with the intention to properly quantifying it. Even, in such 
a simple plot 
a general dependence of ${\rm I}^{\rm TRGB} - {\rm I}^{\rm RC}$ on metallicity 
of the order of 0.21 mag/dex is seen. This is in  good agreement with the 
metallicity dependence of the I-band magnitude of the red clump found by 
Udalski (2000) and Pietrzynski, Gieren and Udalski (2003). However, the large scatter
of the data points around a simple linear relation
indicates that other factors (e.g. dependence on age, a more complicated non-linear
dependence on metallicity) likely 
play an important role.

It is also worthwhile to mention that, as can be seen 
from Fig 2, selecting different subsamples of galaxies and plotting 
${\rm I}^{\rm TRGB} - {\rm I}^{\rm RC}$ versus metallicity one can easily 
obtain very different conclusions about the populational effect on the 
I-band magnitude of red clump stars, from no dependence at all on metallicity to a
very strong dependence. In particular the ${\rm I}^{\rm TRGB} - {\rm I}^{\rm RC}$
is close to 3.6 mag (with a scatter of about 0.1 mag) for all galaxies having 
a mean metallicity larger 
than about -1.9 dex (except DDO 74). In more metal-poor galaxies, the absolute I magnitude 
of the red clump seems to become brighter by 0.4 mag when going from -1.9 to -2.4 dex.
As said before, the current data in Fig. 2 do suggest a non-linear relationship
between the difference of red clump and TRGB absolute magnitudes in the I band, and
the metallicity of these populations.

An  analysis of the population effects on the V band magnitude of red clump stars
can also be done from our data. However in order to compare 
the ${\rm V}^{\rm RC}$,
with the ${\rm I}^{\rm TRGB}$ in a given galaxy we need to adopt a reddening 
for all studied galaxies and  correct the corresponding magnitudes, which evidently
complicates the analysis. Adopting the reddenings
calculated from the maps of Schlegel et al. (1999) and assuming the reddening law 
given in the same 
paper we calculated the ${\rm V}^{\rm RC}_{0} - {\rm I}^{\rm TRGB}_{0}$ and plotted 
it against the metallicity in Figure 3. The diagram is very similar to Fig. 2, and
again the large observed scatter among the data points indicates the presence 
of very significant population effects on the V band magnitudes of red clump stars. 
The total variation of the difference between the red clump V-band magnitude and 
the I-band magnitude of the TRGB over the metallicity range covered by the present
target galaxies appears even larger than in Fig. 2, suggesting an even stronger
population effect on the V absolute magnitudes of red clump stars than in I, which
is supported by the model predictions of Salaris and Girardi.

\section{Summary and conclusions}
Using deep and homogeneous HST photometry we have calculated precise mean I and V band 
magnitudes of the red clump, and have determined the intermediate-population mean
metallicities for some 15 Local Group galaxies. Supplementing these results with 
data from the literature for another 8 galaxies we have compared 
the mean I- and V-band red clump  
magnitudes with the corresponding TRGB I-band magnitudes. Our analysis clearly shows 
that there are very significant population effects on both V and I band magnitudes 
of red clump stars, which fully confirm the results of Girardi and 
Salaris (2001) obtained  from synthetic populations of red clump stars. 
Therefore, in order to use optical photometry of 
these stars for distance determinations to nearby galaxies the population effects
should be very carefully addressed. Since there is at present no way to 
accurately determine 
such  population effects empirically, the only possibility for measuring distances 
from this method is to assume star formation and chemical evolution histories 
in a given galaxy and to calculate, as originally proposed by Girardi and Salaris 
(1999), population corrections for a given band. 
Unfortunately, due to many assumptions (star formation history, chemical evolution history, 
stellar models) the accuracy of such corrections 
(about 0.1 mag) strongly limits the application of optical photometry 
of red clump stars for a precise determination of local distances. Luckily,
as discussed previously in this paper, the situation is much better with using
near-infrared photometry of red clump stars which allow a much more accurate
distance determination to galaxies due to the decreased effect of population
differences on the red clump absolute magnitude. This tendency is of course also
supported by the decreased effect of reddening on the measured mean magnitude,
particularly in the K band.

\acknowledgments
We gratefully acknowledge financial support for this
work from the Chilean Center for Astrophysics FONDAP 15010003, and from
the BASAL Centro de Astrofisica y Tecnologias Afines (CATA) PFB-06/2007. 
Support from the Polish grant N203 387337 and the FOCUS
subsidy of the Fundation for Polish Science (FNP) is also acknowledged.
We would also like to thank the anonymous referee for his suggestions.

\begin{figure}[p]
\vspace*{18cm}
\includegraphics{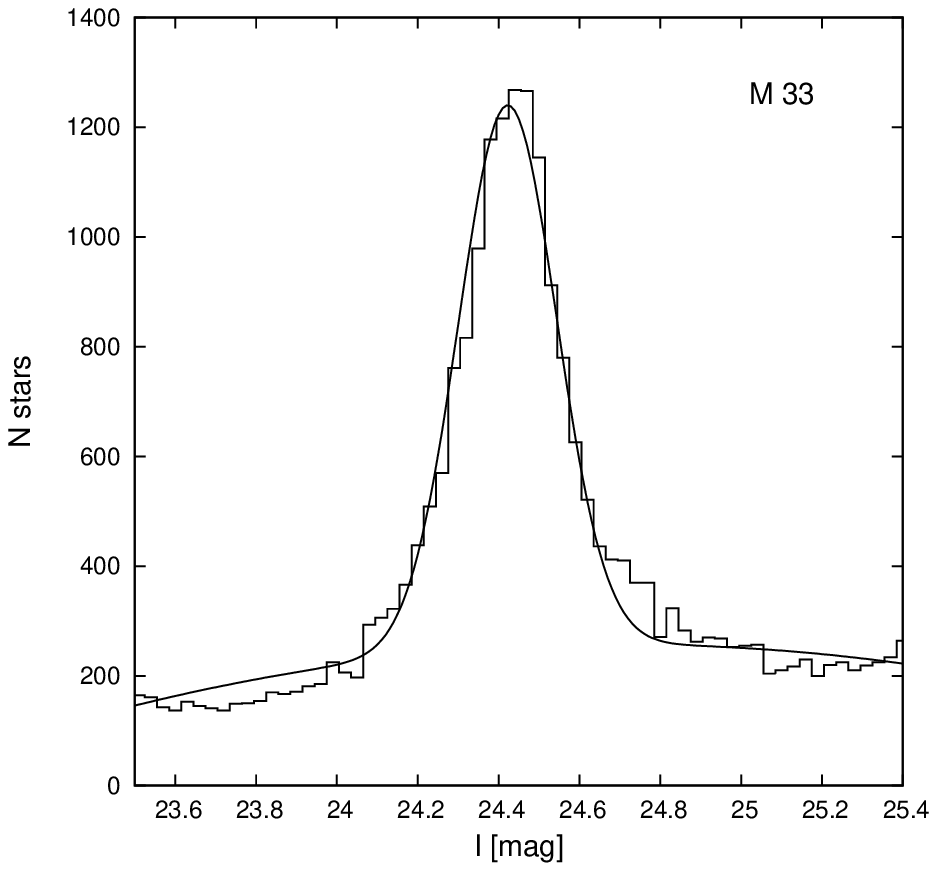}
\includegraphics{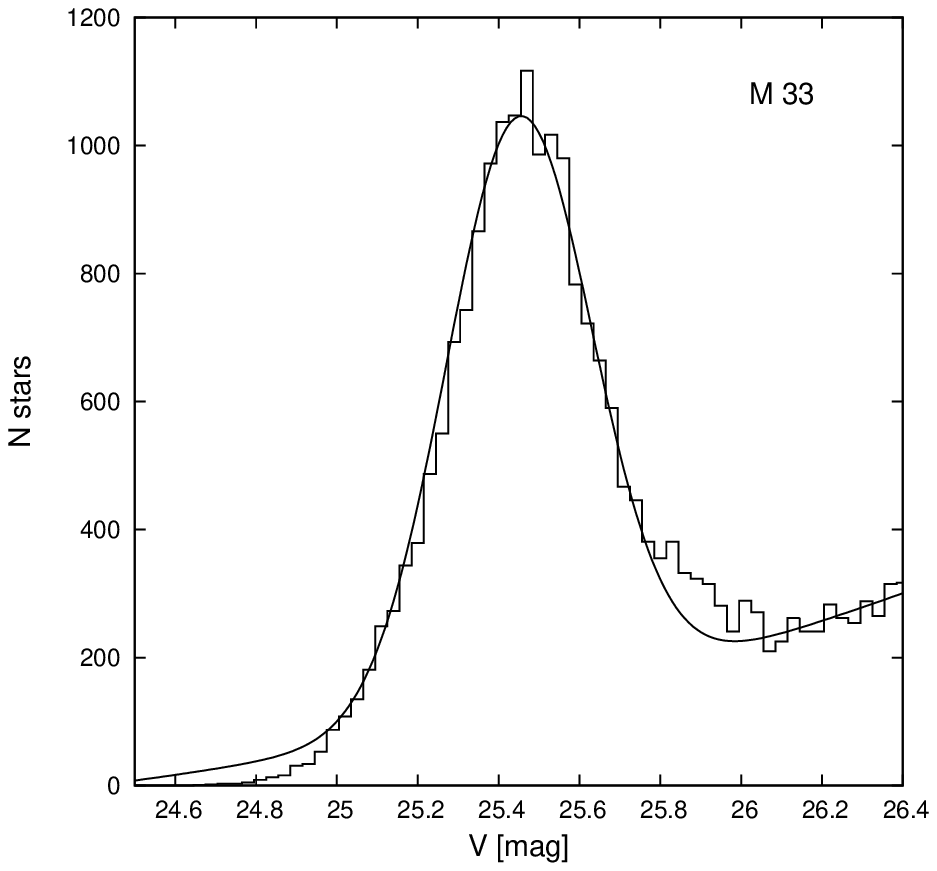}
\includegraphics{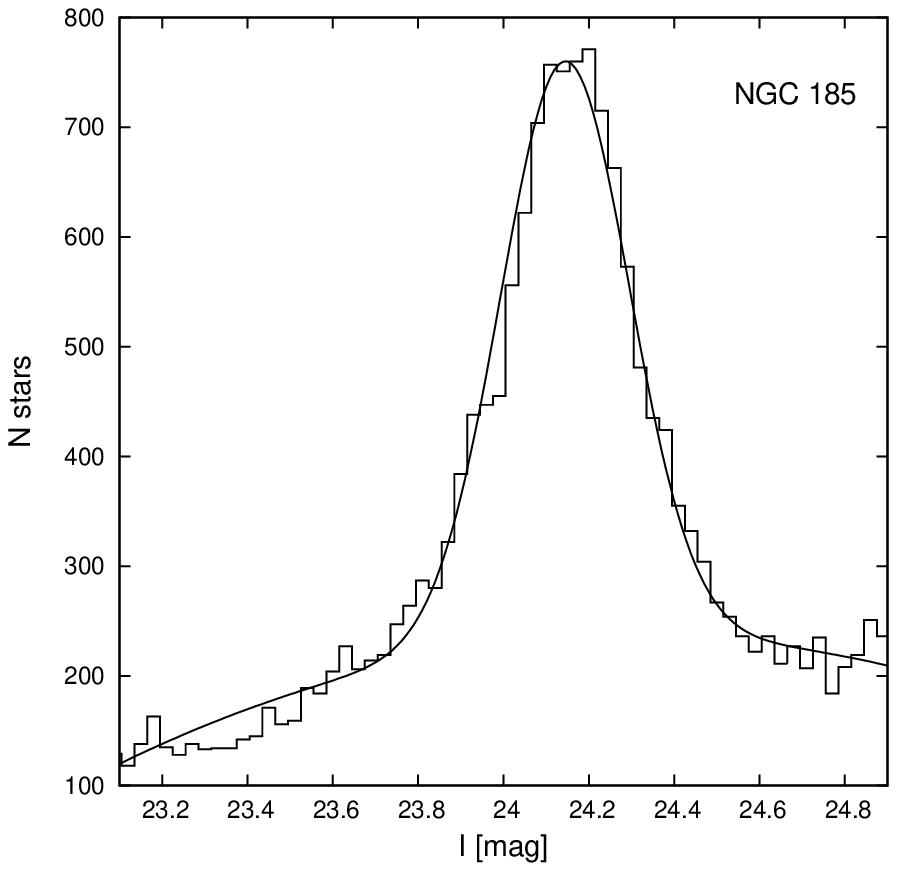}
\includegraphics{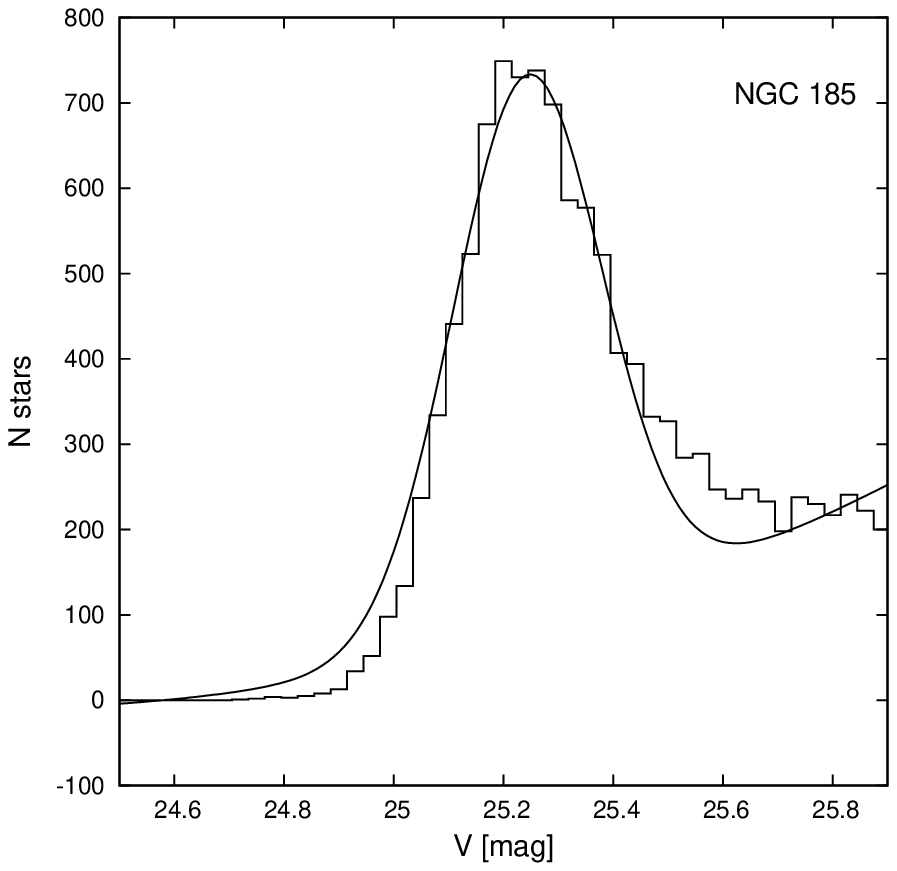}
\caption{Exemplary Gaussian and polynomial fits, according to equation (1) 
to the HST data for nearby galaxies (Jacobs et al. 2009). The excellent accuracy 
of the determination of the mean V and I band magnitudes of red clump stars from 
these data is demonstrated.
}
\end{figure}

%\begin{figure}[p] 
%\vspace*{18cm}
%\special{psfile=wyn.ps  hoffset=-20  voffset=0 hscale=60 vscale=60} 
%\caption{
%}
%\end{figure}  

\begin{figure}[p]
\vspace*{18cm}
\includegraphics{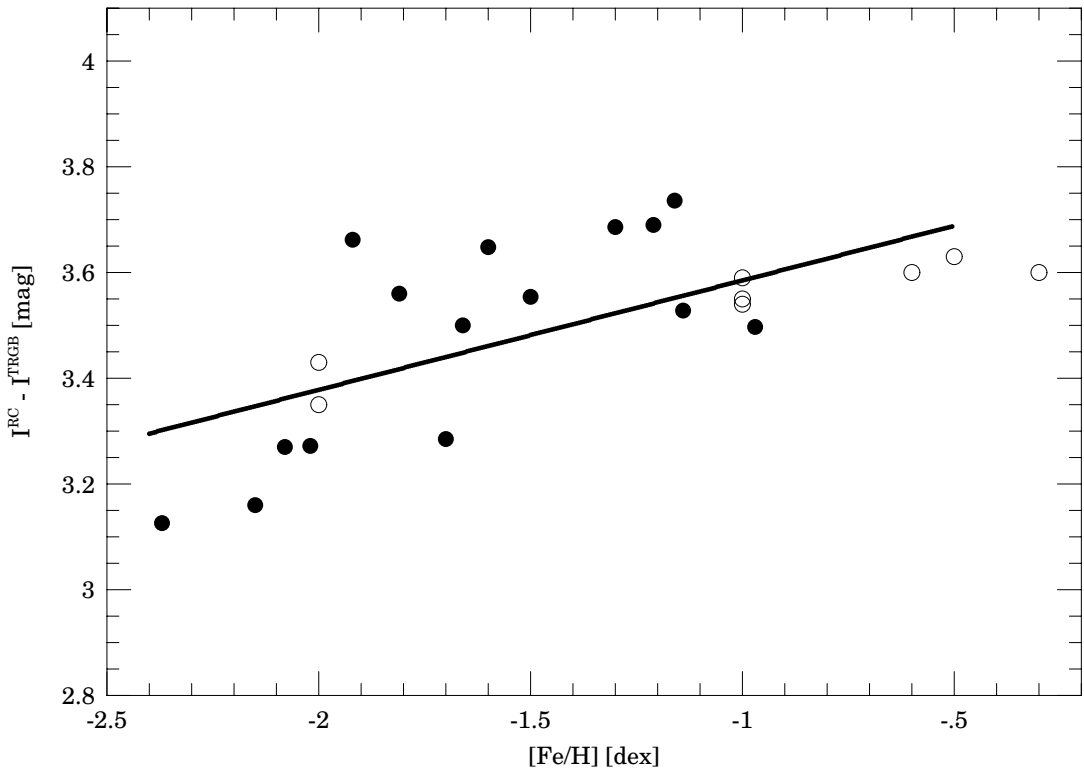}
\caption{The difference between the mean I-band magnitude of red clump stars 
and the I-band TRGB brightness  plotted against the mean metallicity
for these stars, for 23 nearby galaxies.
Filled and open circles correspond to the data obtained in this paper
from recently published homegenous HST photometry, and 
data tabulated by Udalski et al. (2000), respectively. 
Error bars to the magnitudes (not shown in the Figure) are of the 
order of 0.05 mag. 
}
\end{figure}

\begin{figure}[htb]
\vspace*{18cm}
\includegraphics{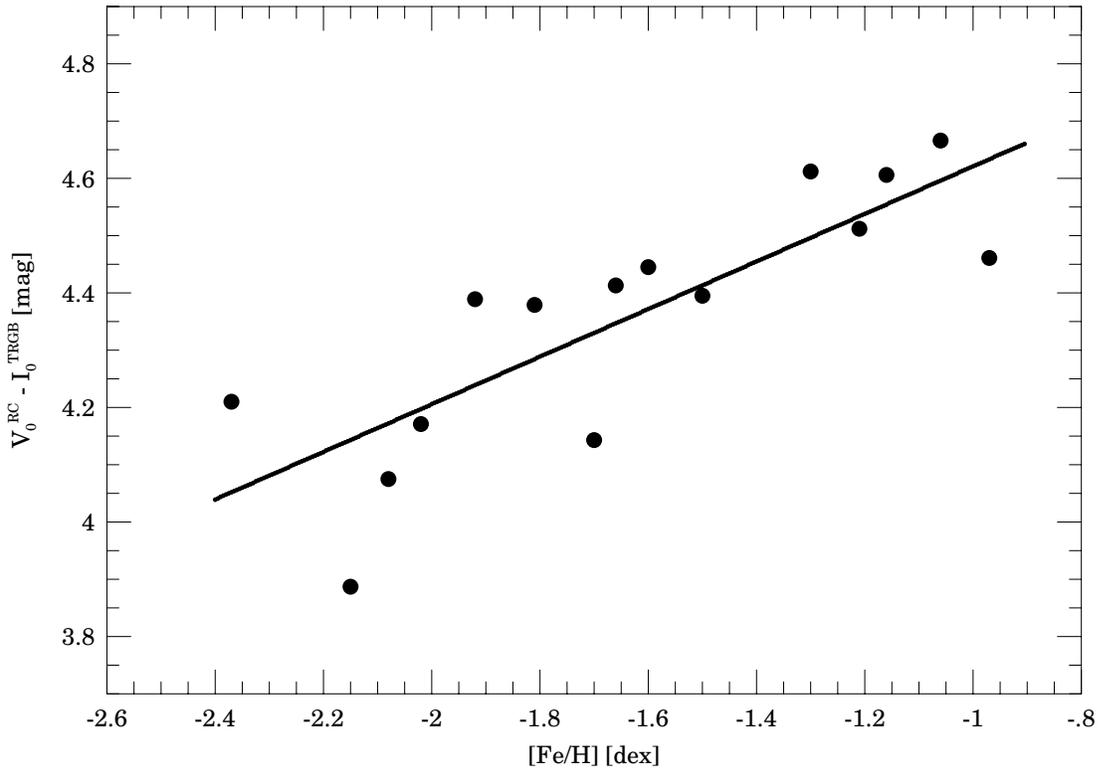}
\caption{
The difference between the mean, absorption-corrected V-band magnitude of red clump stars
and the absorption-corrected I-band magnitude of the TRGB plotted against the
metallicity of the stars, for 15 nearby galaxies. HST data were used for the determination
of the magnitudes.
Error bars to the magnitudes (not shown in the Figure) are of the
order of 0.05 mag.
}
\end{figure}

\clearpage
\begin{deluxetable}{c c c c c c c c c}
%\rotate
\tablewidth{0pc}
\tablecaption{Red clump and TRGB magnitudes, and metallicities calculated from 
homogenous HST data for 15 galaxies}
\tablehead{ \colhead{Galaxy} & \colhead{ ${\rm I}_{\rm RC}$ } & \colhead{${\sigma}_{\rm I}$}
& \colhead{ ${\rm V}_{\rm RC}$ } 
& \colhead{${\sigma}_{\rm V}$} & \colhead{ ${\rm I}_{\rm TRGB}$ } & \colhead{[Fe/H]} 
& \colhead{ ${\sigma}_{[Fe/H]}$ } & \colhead{E(B-V)}
}
\startdata
NGC6822 &  23.478 &    0.004 &   24.907 &   0.004 &   19.950 &   -1.06 &    0.03 &    0.23 \\ 
IC1613 &  23.924 &    0.007 &   24.797 &   0.005 &   20.370 &   -1.50 &    0.02 &    0.03 \\ 
DDO74 &  21.375 &    0.010 &   22.279 &   0.006 &   18.090 &   -1.70 &    0.06 &    0.04 \\ 
Phoenix &  22.660 &    0.024 &   23.593 &   0.010 &   19.160 &   -1.66 &    0.04 &    0.02 \\ 
NGC185 &  24.136 &    0.003 &   25.238 &   0.004 &   20.400 &   -1.16 &    0.03 &    0.18 \\ 
NGC147 &  24.360 &    0.003 &   25.403 &   0.002 &   20.670 &   -1.21 &    0.03 &    0.17 \\ 
DDO69 &  24.002 &    0.015 &   24.927 &   0.007 &   20.730 &   -2.02 &    0.08 &    0.02 \\ 
Pegasus &  24.676 &    0.006 &   25.688 &   0.005 &   20.990 &   -1.30 &    0.03 &    0.07 \\ 
DDO75 &  24.940 &    0.008 &   25.722 &   0.011 &   21.780 &   -2.15 &    0.04 &    0.04 \\ 
M33 &  24.417 &    0.003 &   25.446 &   0.003 &   20.920 &   -0.97 &    0.03 &    0.05 \\ 
E594-004 &  24.436 &    0.013 &   25.680 &   0.016 &   21.310 &   -2.37 &    0.06 &    0.12 \\ 
Antlia &  25.382 &    0.007 &   26.210 &   0.010 &   21.720 &   -1.92 &    0.01 &    0.08 \\ 
E410-005 &  25.940 &    0.007 &   26.777 &   0.005 &   22.380 &   -1.81 &    0.02 &    0.01 \\ 
UGC9128 &  26.020 &    0.009 &   26.854 &   0.010 &   22.750 &   -2.08 &    0.03 &    0.02 \\ 
E294-010 &  26.118 &    0.013 &   26.923 &   0.010 &   22.470 &   -1.60 &    0.02 &    0.01 \\ 
\enddata
\end{deluxetable}

\end{document}